\documentclass[a4paper, amsfonts, amssymb, amsmath, preprint, showkeys, nofootinbib, twoside]{revtex4-1}
\usepackage[english]{babel}
\usepackage{tabularx}
\usepackage[utf8]{inputenc}
\usepackage[colorinlistoftodos, color=green!40, prependcaption]{todonotes}
\usepackage[pdftex, pdftitle={Article}, pdfauthor={Author}]{hyperref} 
\usepackage[version=4]{mhchem}
\usepackage{hyperref}
\usepackage{gensymb}
\begin{document}
\title{Reply to “Comment on ‘Topography of Fermi arcs in t-PtBi$_2$ using high-resolution
angle-resolved photoemission spectroscopy”’
}
\author{Adam Kaminski$^{1, 2}$}
\email{adamkam@ameslab.gov}
\author{Paul C. Canfield$^{1, 2}$}
\author{Benjamin Schrunk$^{2}$}
  \affiliation{$^1$Department of Physics and Astronomy, Iowa State University, Ames, Iowa 50011, USA}
  \affiliation{$^2$Division of Materials Science and Engineering, Ames Laboratory,
US DOE, Ames, Iowa 50011, USA. \\
}

\date{\today} 

\begin{abstract}
In a recent work \cite{Evanoriginal} we demonstrated that the Fermi arcs in t-PtBi$_2$ are indeed disconnected segments of the Fermi surface. We showed that the surface state band crosses E$_F$ between Weyl points and is has finite band gap at all other momenta until it merges with bulk bands. We did not find evidence of a superconducting gap in the spectra down to 3~K. In recent Comment \cite{Boriscomm}, Sergey Borisenko re-analyzed data from repository posted with our paper and concluded that a superconducting gap with i-wave symmetry is present in our data up to temperatures of 19~K. We argue that this claim arises from incorrect identification of the location of Fermi arc and confusion between band gap and superconducting gap. This was compounded by the fact that Author of the Comment analyzed data from interpolated images used for mapping the Fermi surface that are not suitable for extraction of energy gap on meV scale. This fact was previously communicated to the Author of the Comment during e-mail exchange. We maintain that our conclusion of absence of a superconducting gap above 3 K are valid and consistent with recent STM measurements \cite{STM1} performed on the same samples, which found surface superconductivity with T$_c$ = 3~K, below the temperature at which the data for our ARPES paper was measured. 

\end{abstract}

\maketitle

Fermi arcs are unusual objects. Conventionally, Fermi surfaces are closed contours in the momentum space. In Weyl semimetals, Fermi arcs span two Weyl points W+ and W-. The surface band forming them, crosses E$_F$ along arc in momentum space connecting the W+ and W- points, while this band is gapped at other parts of the Brillouin zone, and merges with the bulk band without crossing E$_F$ or reaching the zone boundary.  Pairing and superconductivity has largest effect on the electronic structure  along the Fermi arc because this is the only boundary between occupied and unoccupied states. At other locations, the surface state band is already gapped and therefore shift of the EDC's leading does not directly measure the value of the superconducting gap, but superposition of the two gaps. In fact, the superconducting contribution to leading edge shift should decrease with increasing band gap.

The Author of the Comment performed an analysis of our data mostly outside of the Fermi arcs, thus mistakenly interpreting the band gap as superconducting gap. In Fig. 1a of the comment [2], only points marked N, 1 and 3 are located on the Fermi arc. The peaks in the EDCs for all these three points are located at almost the same value of -3meV, as seen rightmost panel of Fig. 1c of the Comment \cite{Boriscomm} and leading edge enrgies are consistent with absence of a gap.  All other points are measured outside of the tips of the Fermi arc where the band gap is present. These points might be influenced only slightly by the superconducting gap if the measurements were done below T$_c$, which they were not. This renders most of the remaining re-analysis presented in other panels of Fig. 1 and Fig. 2 of the Comment \cite{Boriscomm} irrelevant, as these data are not measured at k$_f$. Demonstrating presence of the band gap outside of the Fermi arcs was indeed one of key points of our paper \cite{Evanoriginal} as shown in Figs. 3 and 4. 

It should be noted that the data plotted in Fig. 3 of the Comment \cite{Boriscomm}, was extracted from interpolated images and was not corrected for analyzer energy calibration, unlike the data presented as EDC’s in our original paper. The data shown in Fig. S1 of the Comment illustrates the danger of analyzing interpolated, uncalibrated image data used for Fermi surface mapping for analysis of superconducting gap on meV scale. We used straight slits in the DA30 Scienta analyzer for measurement to maintain proper momentum mapping. This results in ''smile” shape of the energy profile, where the constant energy line is curved across the detector, with largest deviation near its edges. Software correction does reasonable job close to center of the detector and there is only a small variation in location of E$_F$ at the center in Fig. S1 of the Comment. This becomes worse further from the center and is quite large near the edges where the analysis of the arc data presented in the Comment was performed. This is evident, as some of the leading edges in top panel of Fig. S1 of the comment are located well above average of location of the leading edge the background, with typical errors exceeding 2 meV. For example, red, violet and blue curves are ~3 meV above the average at the center, which would imply negative binding energy. The yellow, green, blue, light green and violet curves show non-monotonic variations between dip and peak near the edges of the detector on both sides within $\sim$ 2 meV. This does not affect the images of the Fermi surface, which are integrated within few meV, but these data are unsuitable for studies of the superconducting gap on meV scale as attempted by the Author of the Comment.

To clarify the location of the Fermi arcs, we plot in Fig. 1, the ARPES intensity at E$_F$, EDCs along relevant cuts and energy values for EDC leading edges along selected cuts. This is actual data that was not interpolated to improve appearance of the image. The E$_F$ is located above 50\% of the leading edges intensity for red EDCs, signifying that the band crosses E$_F$ and energy gap is absent along the Fermi arc, as expected. Outside of the tips of the arcs, the peaks are located away from the E$_F$, consistent with a standard band gap. The leading edge shifts for these EDC are plotted in panel c of Fig. 1. The gap is closed all along the whole arc, with band gap present outside of the tips of the arcs. This contradicts the i-wave scenario, where only a single momentum point – node should have zero gap.
The samples from the same growth as used in our ARPES studies were independently measured by STM [3]. These data are plotted in Fig. 2 and show robust surface superconducting gap that opens only below Tc of ~3`K, therefore could not be observed in our ARPES data which was measured only above that temperature. We note that the DOS has flat bottom within the gap which is consistent of isotropic s-wave symmetry of the order parameter. For i-wave symmetry one would expect V-shaped DOS within the gap. Similar STM data was also recently published \cite{STM2}
using independently grown samples.

It should be noted that one of the current authors (AK) submitted a comment to Nature on August 28, 2026 demonstrating that the spectra presented in \cite{iwave} lacked the signatures of Bogoliubov quasiparticles, which are necessary to establish presence of superconducting gap in the ARPES spectra.

\bibliography{Refs}

\newpage 

\begin{figure}
    \includegraphics[scale=.5]{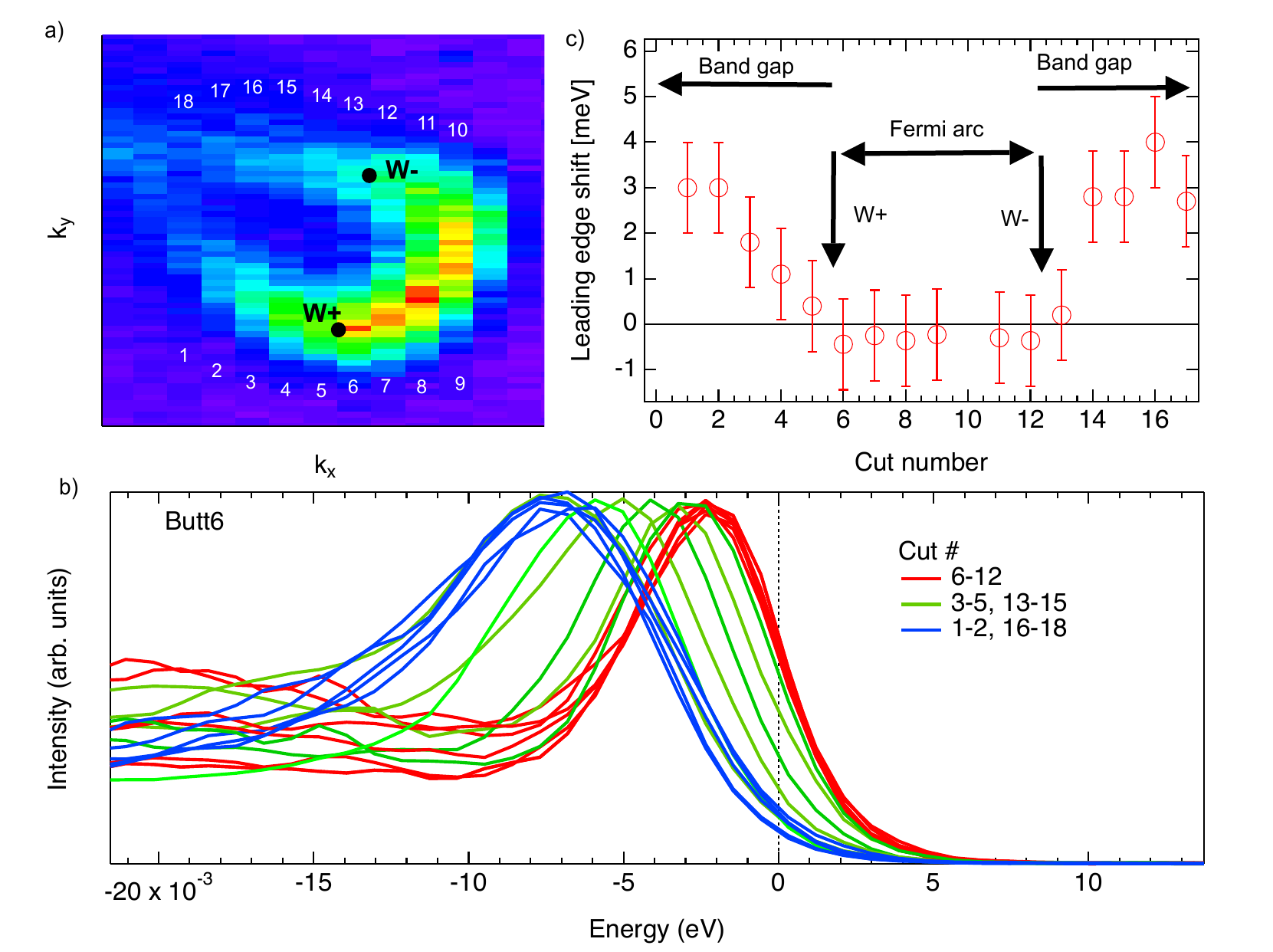}
    \caption{ARPES data from t-PtBi2, in the arc area measured at T=4 K, sample surface superconducting T$_c$ from STM measurements is 3 K. a) ARPES intensity at E$_F$ in the Fermi arc region of momentum space. b) EDCs at for cuts marked in (a) at momentum, where leading edge is closest to E$_F$. Numbers 1- 9 correspond to the lower branch and 10-18 for the upper branch.  c) energy shift from E$_F$ of the leading edge for EDCs shown in (b) indexed by the cut number shown in (a).}
    \end{figure}
    
 \begin{figure}
    \includegraphics[scale=1.5]{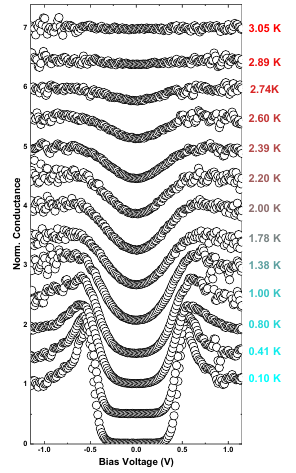}
    \caption{DOS measured by STM on the samples of t-PtBi2 from the same batch as ARPES data, kindly provided by authors of Ref. \cite{STM1}}.

\end{figure}

\end{document}